\documentclass{ws-procs10x7}
\usepackage{balance}

\newcolumntype{d}[1]{D{.}{.}{#1}}

\def\Journal#1#2#3#4{{\it #1} {\bf #2}, #3 (#4)}

\makeindex
\begin{document}

\title{QUANTUM NERNST EFFECT}

\author{HIROAKI NAKAMURA}

\address{Theory and Computer Simulation Center,
National Institute for Fusion Science,\\
Oroshi-cho, Toki, Gifu 509-5292, Japan\\
E-mail: nakamura@tcsc.nifs.ac.jp}

\author{NAOMICHI HATANO}

\address{Institute of Industrial Science,
University of Tokyo,
Komaba, Meguro, Tokyo 153-8505, Japan\\
E-mail: hatano@iis.u-tokyo.ac.jp}

\author{RY\=OEN SHIRASAKI}

\address{Department of Physics, Yokohama National University,\\
Tokiwadai, Hodogaya-ku, Yokohama 240-8501, Japan\\
E-mail: sirasaki@phys.ynu.ac.jp}


\twocolumn[\maketitle\abstract{
We report our recent predictions on the quantum Nernst effect, a novel thermomagnetic effect in the quantum Hall regime.
We assume that, when the chemical potential is located between a pair of neighboring Landau levels, edge currents convect around the system.
This yields theoretical predictions that the Nernst coefficient is strongly suppressed and the thermal conductance is quantized.
The present system is a physical realization of the non-equilibrium steady state.}
]

\section{Introduction}

The adiabatic Nernst effect arises in a conductor bar under a magnetic field $B$ in the $z$ direction and a temperature bias $\Delta T$ in the $x$ direction.
The conductor is electrically and thermally insulated on all surfaces, except that heat baths are attached to the surfaces facing the $+x$ and $-x$ directions, which produces the temperature bias.
A classical-mechanical consideration gives the following:
electrons carrying the heat current in the $x$ direction are deflected to the $y$ direction because of the Lorentz force generated by the magnetic field in the $z$ direction, and thereby produce a voltage difference (the Nernst voltage) $V_\mathrm{N}$ in the $y$ direction.
The Nernst coefficient is defined by
\begin{equation}\label{eq1}
N\equiv -\frac{L}{W} \frac{V_\mathrm{N}}{B \Delta T},
\end{equation}
where $W$ and $L$ are the width and the length of the conductor bar, respectively.
We define the temperature bias such that $\Delta T>0$ if the temperature is higher in the heat bath on the $-x$ surface than that on the $+x$ surface.
We also define the Nernst voltage such that $V_\mathrm{N}>0$ if the voltage is higher on the $+y$ surface than on the $-y$ surface.
(We always put $B>0$ here and hereafter.)
The above classical-mechanical consideration, where no scattering is taken into account, gives a positive Nernst coefficient.
In fact, electron scattering can make the Nernst coefficient both positive and negative.\cite{Nakamura99}

In a recent article,\cite{NHS05} we considered the Nernst effect in the quantum Hall regime, that is, the Nernst effect of the two-dimensional electron gas in semiconductor heterojunctions under a strong magnetic field (namely a Hall bar) at low temperatures, low enough for the mean free path to be greater than the system size.
We theoretically predicted that, when the chemical potential is located between a pair of neighboring Landau levels:
\begin{enumerate}
\renewcommand{\labelenumi}{(\roman{enumi})}
\item the Nernst coefficient is strongly suppressed;
\item the thermal conductance in the $x$ direction is quantized.
\end{enumerate}
Hasegawa and Machida are planning an experiment\cite{nernst.experiment} in the setup of the present theory.
In what follows, we review our predictions and numerical demonstration.

\section{Convection of edge currents}

\begin{figure}[b]
\begin{center}
\includegraphics[width=0.9\columnwidth,clip]{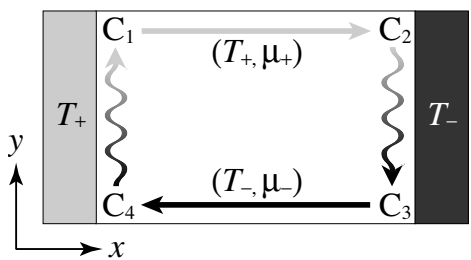}
\end{center}
\caption{A schematic view of the convection of an edge current in a Hall bar under the setup for the Nernst effect.}
\label{fig.idea}
\end{figure}

Our argument is based on a simple assumption that edge currents\cite{Halperin82} convect around the system.
We here explain our idea (Fig.~\ref{fig.idea}).
Since the Hall bar is electrically insulated, edge currents circulate along the edges of the Hall bar when the chemical potential is between a pair of neighboring Landau levels.
An edge current on the left end of the Hall bar is, while running from the corner C$_4$ to the corner C$_1$, heated up by the heat bath with a temperature $T_+$ and is equilibrated to the Fermi distribution $f(T_+,\mu_+)$ with the temperature $T_+$ and a chemical potential $\mu_+$ around the corner C$_1$.
Since the upper edge is electrically and thermally insulated, the edge current runs ballistically from the corner C$_1$ to the corner C$_2$, maintaining the Fermi distribution $f(T_+, \mu_+)$ all the way.
The edge current, upon arriving at the corner C$_2$, encounters the other heat bath with a temperature $T_-$ and is equilibrated to the Fermi distribution $f(T_-, \mu_-)$, arriving at the corner C$_3$.
The edge current along the lower edge runs ballistically from the corner C$_3$ to the corner C$_4$, maintaining the Fermi distribution $f(T_-, \mu_-)$ all the way.
The Nernst voltage is then given by
\begin{equation}\label{voltage}
V_\mathrm{N}=\frac{\Delta\mu}{\mathrm{e}}\equiv\frac{\mu_+-\mu_-}{\mathrm{e}}
\end{equation}
for the temperature bias $\Delta T\equiv T_+-T_->0$, where $\mathrm{e}(<0)$ is the charge of the electron.

Incidentally, a convecting edge current constitutes the non-equilibrium steady state (NESS), a new concept that attracts much attention in the field of non-equilibrium statistical physics.\cite{Ruelle00,Ogata04}
The non-equilibrium steady state is almost the first statistical state of a quantum system far from equilibrium that can be handled analytically.
It consists of a pair of independent currents running in different directions with different temperatures and different chemical potentials.
In most studies, the non-equilibrium steady state has been considered in a one-dimensional non-interacting electron system and hence has been an almost purely mathematical concept.
The pair of the upper and lower edge currents in Fig.~\ref{fig.idea}, however, can be regarded as a non-equilibrium steady state.
We consider it valuable to give a physical realization to the mathematical-physical concept.

\section{Nernst coefficient and heat conductance}

Let us describe our calculation briefly.
We define the electric current and the heat current in the form
\begin{equation}\label{current}
I_\mathrm{e}\equiv\langle \mathrm{e}v\rangle
\quad\mbox{and}\quad
I_Q\equiv\langle(E-\mu)v\rangle,
\end{equation}
where the thermal average is given by
\begin{equation}\label{average}
\langle A\rangle
\equiv\frac{1}{\pi}\sum_{n=0}^\infty
\int_{-k_\mathrm{m}}^{k_\mathrm{m}}
Af_{n,k}(T(y_k),\mu(y_k))dk.
\end{equation}
with $y_k\equiv\hbar k/|\mathrm{e}|B$.
The subscript $n$ stands for the channel of the edge current.
The integration limits $\pm k_\mathrm{m}$ are the maximum and minimum possible momenta.
The function $f_{n,k}$ is the Fermi distribution at the energy of the state of the $n$th channel with the momentum $k$.
The functions $T(y)$ and $\mu(y)$ denote the temperature and the chemical potential of an edge current running at $y$.
The convection of an edge current shown in Fig.~\ref{fig.idea} gives $(T(y_k),\mu(y_k))=(T_+,\mu_+)$ for the upper edge states and $(T_-,\mu_-)$ for lower edge states.
We can hence express the currents~(\ref{current}) in terms of $(T_+,\mu_+)$ and $(T_-,\mu_-)$, and thereby in terms of the first order of the Taylor expansion with respect to $\Delta T\equiv T_+-T_-$ and $\Delta\mu\equiv\mu_+-\mu_-$.

We then put $I_\mathrm{e}=0$ because the system is electrically insulated.
This condition relates $\Delta\mu$ to $\Delta T$, yielding the Nernst coefficient~(\ref{eq1}) with the Nernst voltage~(\ref{voltage}), or
\begin{equation}\label{eq10}
N=\frac{1}{|\mathrm{e}|B}\frac{L}{W}\frac{\Delta\mu}{\Delta T}.
\end{equation}
Applying the relation between $\Delta\mu$ and $\Delta T$ to the Taylor expansion of the heat current $I_Q$, we obtain the heat conductance
\begin{equation}\label{eq11}
G_Q\equiv \frac{I_Q}{\Delta T}.
\end{equation}
For details of the calculation, refer to the original article, Ref.~\refcite{NHS05}.

\section{Predictions}

We can understand our predictions based on the convection of edge currents.
First, the number of the conduction electrons is conserved during the convection.
Hence the difference in the chemical potential of the upper edge current and that of the lower edge current is of a higher order of the difference in the temperature of the upper and lower edge currents;
that is, $\Delta\mu=o(\Delta T)$.
The Nernst coefficient~(\ref{eq10}) hence vanishes as a linear response, or $N=0$ in the limit $\Delta T\to0$.

Second, the heat current $I_Q$ in the $x$ direction is carried by the ballistic edge currents along the upper and lower edges;
the total heat current is the difference in the heat carried by the upper edge currents and the lower edge currents.
The edge currents are quantized as long as the chemical potential remains between a pair of neighboring Landau levels.
The heat current hence has quantized steps as a function of $B$.
The heat current is $M$ times a unit current when there are $M$ channels of the edge current.
After some algebra, we arrive at the conclusion that the heat conductance $G_Q$ is quantized as
\begin{equation}\label{eq.lowT}
\frac{G_Q}{T}=\frac{\pi{k_\mathrm{B}}^2}{3\hbar}M
\quad (M=1,2,3,\cdots)
\end{equation}
when the chemical potential is located between the $M$th and $(M+1)$th Landau levels.

\section{Numerical demonstration}

We now present a numerical demonstration of our predictions using a confining potential in the $y$ direction in the form\cite{Komiyama92}
\begin{equation}
 V(y) = 
  \left\{
    \begin{array}{ll}
      0           & \mbox{for}\quad |y| \leq \frac{w}{2} , \\
      \frac{m \omega_0^2}{2} \left( |y| - \frac{w}{2} \right)^2
             & \mbox{for}\quad \frac{w}{2}< |y| <\frac{W}{2}.
    \end{array}
  \right.
  \label{v.eq}
\end{equation}
We used the following parameter values:
the effective mass is $m=0.067 m_0$ for GaAs, where $m_0$ is the bare mass of the electron;
the size of the Hall bar is $L=20\mu$m and $W=20\mu$m (less than the mean free path at low temperatures\cite{Tarucha92}) with $w=16\mu$m;
the confining potential is given by $V(\pm W/2) = 5.0$eV, the work function of GaAs;
the chemical potential at equilibrium is $\mu=15$meV, or the carrier density $n_\mathrm{s}=4.24\times 10^{15}\mathrm{m}^{-2}$.

\begin{figure}[t]
\begin{center}
\includegraphics[width=0.9\columnwidth]{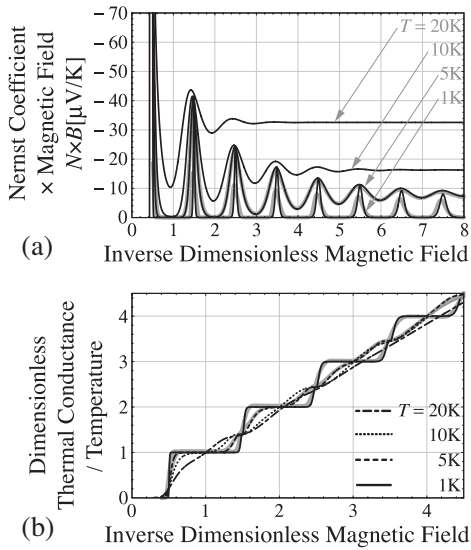}
\end{center}
\caption{Scaling plots of (a) $N \times B$ and (b) $G_Q/T\times3\hbar/\pi{k_\mathrm{B}}^2$, both against $m\mu/\hbar|\mathrm{e}|B$ at $T=1,5,10$ and $20$K for $1\mathrm{T}\leq B\leq 20\mathrm{T}$.
The gray curves in each panel indicate results that take account of impurity scattering at $T=1$ and $5$K; see text.}
\label{results}
\end{figure}

We thus evaluated the Nernst coefficient $N$ and the thermal conductance $G_Q$ as in Fig.~\ref{results}.
Our predictions $N=0$ and Eq.~(\ref{eq.lowT}) are indeed realized at low temperatures and when the chemical potential is located between a pair of neighboring Landau levels.
(The gray curves in Fig.~\ref{results} are our new results of the self-consistent Born approximation in the case where we took account of impurity scattering of strength $\hbar/\tau=1.0\times 10^{-4}$eV.
See Ref.~\refcite{future} for details.)
We also note that the Nernst coefficient is generally negative in the present calculation.

\section{Summary}

We predicted a novel quantum effect of the two-dimensional electron gas, in close analogy to the quantum Hall effect.
When the chemical potential is between a pair of Landau levels, the edge currents suppress the Nernst coefficient and quantize the thermal conductance.
The system is a physical realization of the non-equilibrium steady state.

The precise forms of the peaks and the steps in Fig.~\ref{results} can be different when we take account of electron scattering.
The electronic states extend over the system when the chemical potential is close to a Landau level, namely when $m\mu/\hbar|\mathrm{e}|B=n+\frac{1}{2}$.
Then the heat current is carried mainly by the bulk states and we have to take account of impurities as well as electron interactions possibly.\cite{future}

We comment on other approaches to the quantum Nernst effect.
Kontani derived\cite{Kontani02,Kontani03} on the basis of the Fermi liquid theory, general expressions of the Nernst coefficient and the thermal conductivity of strongly correlated electron systems such as high-$T_\mathrm{c}$ materials.
Akera and Suzuura\cite{Akera04} considered with the use of thermohydrodynamics, the Ettingshausen effect, the reciprocal of the Nernst effect.
The quantum behavior predicted here, however, was not reported in either studies.

\section*{Acknowledgments}
The authors express sincere gratitude to Dr.~Y.~Hasegawa and Dr.~T.~Machida for useful comments on experiments of the Nernst effect and the quantum Hall effect.
This research was partially supported by the Ministry of Education, Culture, Sports, Science and Technology, Grant-in-Aid for Exploratory Research, 2005, No.17654073 as well as by the Murata Science Foundation.
N.H.~gratefully acknowledges the financial support by the Sumitomo Foundation.


\begin{thebibliography}{99} 

\bibitem{Nakamura99}
H. Nakamura, K. Ikeda, Y. Ishikawa, A. Suzuki and H. Shirai, \Journal{Jpn. J. Appl. Phys.}{38}{5745}{1999}.

\bibitem{NHS05}
H. Nakamura, N. Hatano and R. Shirasaki, \Journal{Solid State Commun.}{135}{510}{2005}.

\bibitem{nernst.experiment}
Y. Hasegawa and T. Machida, private communication.

\bibitem{Halperin82}
B. I. Halperin, \Journal{Phys. Rev. B}{25}{2185}{1982}.

\bibitem{Ruelle00}
D. Ruelle, \Journal{J. Stat. Phys.}{98}{57}{2000}.

\bibitem{Ogata04}
Y. Ogata, \Journal{Phys. Rev. E}{66}{016135}{2004}.

\bibitem{Komiyama92}
S. Komiyama, H. Hirai, M. Ohsawa, Y. Matsuda, S. Sasa, and T. Fujii, \Journal{Phys. Rev. B}{45}{11085}{1992}. 

\bibitem{Tarucha92}
S. Tarucha, T. Saku, Y. Hirayama, and Y. Horikoshi, \Journal{Phys. Rev. B}{45}{13465}{1992}.

\bibitem{future}
H. Nakamura, N. Hatano, and R. Shirasaki, in preparation.

\bibitem{Kontani02}
H. Kontani, \Journal{Phys. Rev. Lett.}{89}{237003}{2002}.

\bibitem{Kontani03}
H. Kontani, \Journal{Phys. Rev. B}{67}{014408}{2003}.

\bibitem{Akera04}
H. Akera and H. Suzuura, cond-mat/0409498.

\end{thebibliography}
\end{document}